\documentclass[aps,pre,twocolumn,floats,showpacs]{revtex4}
\usepackage{epsfig}
\usepackage{color}
\usepackage{graphicx}
\usepackage{caption}
\usepackage{amsmath, amssymb}
\usepackage{bbold}
\usepackage{bm}
\usepackage{latexsym}
\usepackage{xcolor}
\usepackage[normalem]{ulem}

\begin{document}
\newcommand{\hide}[1]{}
\newcommand{\tbox}[1]{\mbox{\tiny #1}}
\newcommand{\half}{\mbox{\small $\frac{1}{2}$}}
\newcommand{\sinc}{\mbox{sinc}}
\newcommand{\const}{\mbox{const}}
\newcommand{\trc}{\mbox{Tr}}
\newcommand{\intt}{\int\!\!\!\!\int }
\newcommand{\ointt}{\int\!\!\!\!\int\!\!\!\!\!\circ\ } 
\newcommand{\eexp}{\mbox{e}^}
\newcommand{\bra}{\left\langle}
\newcommand{\ket}{\right\rangle}
\newcommand{\EPS} {\mbox{\LARGE $\epsilon$}}
\newcommand{\ar}{\mathsf r}
\newcommand{\im}{\mbox{Im}}
\newcommand{\re}{\mbox{Re}}
\newcommand{\bmsf}[1]{\bm{\mathsf{#1}}}
\newcommand{\beq}{\begin{equation}}
\newcommand{\eeq}{\end{equation}}
\newcommand{\bea}{\begin{eqnarray}}
\newcommand{\eea}{\end{eqnarray}}
\definecolor{red}{rgb}{1,0.0,0.0}
\title{Complexity in multi-qubit and many-body systems}
\author{Imre Varga}
\affiliation{Department of Theoretical Physics, Institute of Physics, Budapest University of Technology and
Economics, Műegyetem rkp. 3., H-1111 Budapest, Hungary}
\date{\today}
\begin{abstract}
Characterizing complexity and criticality in quantum systems requires diagnostics that are both computationally tractable and physically insightful. We apply a measure of quantum 
state complexity for $n$-qubit systems, defined as the divergence between the Shannon or von Neumann entropy of the computational basis distribution and the second-order Rényi entropy. 
This quantity has already been used earlier termed as structural entropy and it is particularly powerful as the
Rényi entropy is directly related to state purity, linear entropy, and the inverse participation ratio, providing a clear physical grounding. While other Rényi orders could be used, the second order offers 
a deep and established connection to these key physical quantities. We first validate the measure in canonical noise channels, showing it peaks at the boundary between quantum and classical regimes. 
We then demonstrate its power in many-body physics. For systems exhibiting a many-body localization transition — including deformed random matrix ensembles and a disordered Heisenberg spin chain 
— the complexity measure reliably signals the crossover from integrable/localized to quantum-chaotic/ergodic phases. Crucially, the maximum complexity occurs in the non-ergodic yet extended states 
at the transition, precisely capturing the critical region where the system is neither fully localized nor thermalized. Furthermore, within the chaotic phase, the measure correlates with the survival probability 
of local excitations, revealing a spectrum of thermalization properties. Our results establish that the entropic complexity is a simple, versatile, and effective probe for identifying nontrivial quantum regimes 
and transitions giving a new and alternative insight into such systems.
\end{abstract}

\maketitle

\section{Introduction}

The development of controllable multi-qubit systems~\cite{preskill2018quantum} is central to the advancement of quantum computation and quantum information science. While experimental 
platforms~\cite{Kjaergaard2020} such as superconducting circuits, trapped ions, and neutral atoms have demonstrated high-fidelity operations for small numbers of qubits, scaling these systems 
introduces significant challenges. These arise not only from engineering constraints but also from fundamental features of quantum mechanics, particularly the structure of entanglement~\cite{Coffman2000}
and the sensitivity of quantum coherence to environmental noise~\cite{Zurek2003, schlosshauer2007decoherence}

A key constraint is the {\it monogamy of entanglement}, a property of quantum correlations which prevents a single qubit from being maximally entangled with more than one other qubit. 
While this property underlies the security of quantum communication protocols, it complicates the characterization and manipulation of multi-partite entangled states in quantum processors. 
The exponential growth of the Hilbert space $2^n$ with $n$ qubits leads to a combinatorial explosion in the possible entanglement configurations, making it difficult to classify 
or even measure entanglement efficiently for large systems~\cite{Coffman2000}.

In parallel, decoherence remains a major limiting factor in practical quantum computing. While noise processes such as dephasing and amplitude damping are well understood and often mitigated 
for one- and two-qubit systems, they exhibit more complex correlations in multi-qubit settings. In such systems, local errors can spread due to entanglement, and non-Markovian effects may become relevant. 
This sensitivity to decoherence emphasizes the need for robust error correction schemes, which themselves demand precise control and understanding 
of multi-qubit interactions~\cite{preskill2018quantum, schlosshauer2007decoherence}.

Characterizing the quantum state of a many-qubit system is another central challenge. Standard quantum state tomography becomes infeasible beyond $\sim10$ qubits due to the exponential scaling of measurement requirements. 
As a result, alternative approaches such as randomized benchmarking, direct fidelity estimation, and learning-assisted tomography have been developed to provide partial yet meaningful insights into quantum states and 
processes~\cite{elben2020cross, flammia2011direct}.

Given the challenges of describing quantum states in high-dimensional Hilbert spaces, new frameworks have been proposed to capture the complexity~\cite{Lloyd} of quantum states beyond entanglement alone. One such approach 
is based on entropic measures, which quantify how information is distributed across the components of a quantum state. These measures provide a statistical handle on the spread, structure, and randomness of quantum states, 
often linking to concepts from quantum chaos and thermalization.

Among these, the concept of entropic complexity — typically quantified using generalized Shannon or Rényi entropies — has gained attention as a unifying tool to assess how structured or delocalized a quantum state is in a given basis, 
most commonly the computational (Fock) basis. In particular, entropic complexity can reflect a state’s localization in that basis, its sensitivity to perturbations, and its proximity to thermal equilibrium 
\cite{Luitz2015, Varga2023}.

This approach has been successfully applied to simple two-level systems (qubits), both isolated and embedded in larger many-body contexts. In these systems, entropic measures offer insight into the interplay between coherence, 
entanglement, and classical stochasticity. For example, in driven or noisy qubit systems, one can observe transitions from low to high entropic complexity corresponding to dynamical changes such as dephasing, delocalization, 
or quantum chaos~\cite{Deutsch2018, Garrison2018, Varga2025}.

The entropic perspective naturally connects to theoretical models of quantum thermalization. Deformed random matrix theory (RMT), including the Two-Body Random Interaction Ensemble (TBRE) and its variants, provide tractable models 
for studying how localized excitations in many-body systems spread and eventually relax under unitary evolution~\cite{Feingold1986, Benet2003}. These models emphasize the role of interaction structure and entanglement in shaping 
the survival probability of an initial state and the long-time behavior of closed quantum systems.

In many-body localized (MBL) systems, where disorder prevents full thermalization, entropic complexity remains bounded and the eigenstates remain localized in Fock space. This is in contrast to chaotic systems, where entropic 
measures approach their maximal values in line with the predictions of the eigenstate thermalization hypothesis (ETH). Entropic complexity thus serves as a powerful diagnostic tool for understanding the transition from integrability 
to chaos and for characterizing the structure of many-body quantum states~\cite{Luitz2015, Varga2025}.

This article aims to explore these ideas in depth, focusing on the entropic structure of multi-qubit states and how it reflects the interplay between entanglement constraints, decoherence, and emergent collective behavior in realistic quantum 
computing architectures.

More precisely the entropic complexity of $n$-qubit systems is investigated using differences of the Shannon-entropy and the R\'enyi entropy of order 2 that has proved to be an interesting and effective quantity to characterize states 
as a function of some parameter, running over an interval between two extremes with essentially trivial understanding. The parameter dependence of this quantity can give insight in a mixed situation. Hence the system can be defined 
as an interpolation between two extreme cases, where our complexity measure vanishes and at an intermediate value the maximum complexity reflects an important case with outstanding properties. The state with maximal complexity 
still keeps the properties from one extreme but already shows properties of the other extreme case and hence reflects a certain type of turning point, cross-over between the two extreme cases. Hereby first we derive the basic formulas 
applied in the case when most generally a quantum state can be represented by a density matrix. Then in the next chapter we apply it to describe multi-qubit states first in the presence of decoherence due to noisy environment and then 
in the presence of dephasing due to the interaction with the environment. Then we turn towards the more general problem of many-body systems using standard RMT modelling and also based on the TBRE. The third problem we discuss 
there is the many-body localization transition in relatively short Heisenberg spin-chains under the effect of local, random magnetic field. Finally we also apply the entropic complexity concept charaterizing the survival probability in many-body 
and complex systems. Conclusions are left for the last chapter.

\section{The derivation of the complexity based on the density matrix}
As the most general, and at the same time basis independent description of a quantum system is based on its density matrix, $\rho$, that has by definition a unit trace,
$\trc\left \{\rho\right\}=1$. The most important quantities to be invoked for our purposes are the von Neumann or Shannon entropy,
\beq
S=-\trc \left \{\rho \ln \rho \right \}
\label{entropy}
\eeq
together with an appropriate generalization, the special R\'enyi entropy~\cite{Renyi} of order 2 that is directly connected to purity and the so-called IPR, the inverse participation ratio which is given as
\beq
R_2=-\ln \trc \left \{\rho^2\right\}.
\label{purity}
\eeq
The parameter that we will calculate will be termed as entropic complexity, i.e. 
$S_C$ defined using definitions Eqs.~(\ref{entropy}, \ref{purity}) as
\beq
S_C = S - R_2 = -\trc \left \{\rho \ln \rho \right \} + \ln \trc \left \{\rho^2\right\}.
\label{sstr}
\eeq
Since these quantities are all calculated as traces of several combinations of the density matrix, therefore the calculation reduces to a diagonalization of $\rho$ which can be calculated
as the appropriate sums over the eigenvalues.

The entropic complexity we use here is a resurrection of the structural entropy that has been successfuly  used in the past characterizing distributions.~\cite{VargaPipek}
The main reason for choosing $S$ and $R_2$ is practical rather than philosophical: both of them are extremely well studied, widely used in quantum information, and experimentally accessible. 
$S$ measures overall mixedness and is standard in discussions of entanglement, distillability, decoherence, and usefulness of noisy entangled states. $R_2$, through $\trc\{\rho^2\}$, quantifies purity 
and therefore how concentrated or spiky the state’s spectrum is; it’s also relatively easy to estimate in the lab. 
Their difference, $S_C=S-R_2$, is then used as a diagnostic tool: it captures how different the “global uncertainty” view (from $S$) is from the “effective purity” view (from $R_2$). Importantly, a reasonable notion 
of complexity should satisfy some basic criteria~\cite{Lloyd, LMC}: ($i$) it should be nonnegative, ($ii$) it should be straightforward to evaluate or measure, and ($iii$) when you vary a natural external, e.g. 
noise parameter $p\in [0,1]$, it should behave trivially at the extremes. For instance, in the cases analyzed in details, the state is either fully coherent and almost pure at one end of the interval, 
or fully classical/fully mixed at the other end; both of those are operationally simple, so we want the complexity measure to vanish there. $S_C=S-R_2$ does exactly that: it is zero at both extremes and becomes 
positive in between, reaching a maximum at some intermediate $p^*$. We interpret that peak as the regime where the state is neither perfectly ordered nor fully random, but instead most structured — noisy, 
yet still carrying quantum correlations in a useful, distillable, or activatable way. It is true that nothing forces us to use $S-R_2$  specifically; one can define analogous families such as $S-R_m$  for higher 
Rényi orders $m>1$, or even spectral spread measures like $R_m-R_{m+1}$. These also tend to be nonnegative, vanish at the trivial endpoints, and peak in the interior. They shift the peak position and can probe 
robustness of that most complex window under stronger noise. However, beyond $m=2$ there is much less accumulated experience: higher-order Rényi entropies are harder to estimate experimentally, and their 
operational meaning (e.g. for teleportation benchmarks, distillation thresholds, or hidden non-locality activation) is less standardized. So $S_C=S-R_2$  is not uniquely fundamental — it’s just the cleanest, most interpretable, 
and most mature representative of this general class of complexity diagnostics that vanish at the extremes and has a maximum in between. 

Hereby we use  $S_C$ for the investigation of entangled quantum systems that are the essential models of quantum computing and quantum communication. We investigate its applicability 
in the case when a noisy channel and environment gradually destroys the original quantum coherence, quantumness and entanglement and eventually produce a classically mixed state. For this purpose, we have 
to mention our recent work~\cite{Varga2025} on the investigation of single two-level systems (TLS) also termed as qubits that serves as a basic building block to introduce a parameter, statistical entropic complexity 
measure~\cite{Varga2023} which has been successful for many other studies~\cite{VargaPipek} in the past. We have to mention that R\'enyi entropies have already been applied as possible generalizations of 
measures of complexity~\cite{Renyi2}.

 It is also very similar to the so-called L\'opez-Ruiz-Mancini-Calbet (LMC) complexity~\cite{LMC} parameter but is well-founded and has roots 
back to localization properties and hence $S_C$ has been useful as $S_{str}$ rather describing the shape of various probability distribution functions (PDF-s). It is a non-negative quantity and the more the PDF deviates from 
a uniform distribution the larger it becomes, hence its usage describing the shape of a PDF. Indeed the LMC parameter and our $S_{str}$ have been shown to be practically equivalent~\cite{Lopez, Varga2023}.

\section{Multi-qubit state in noise induced decoherence}

The understanding of a multi-qubit quantum computer in an arguably noisy environment is an essential problem even nowadays. In order to investigate the vulnerability of these systems it can be modelled~\cite{NielsenChuang} 
using the so called Werner states~\cite{Werner} originally introduced for 2-qubit states. In this chapter we wish to introduce and investigate the behavior of its straightforward generalization for the case of $n$ qubits where $n=1,2,3,...$. 
\begin{figure}
\begin{center}
\epsfxsize=8.4cm
\leavevmode
\epsffile{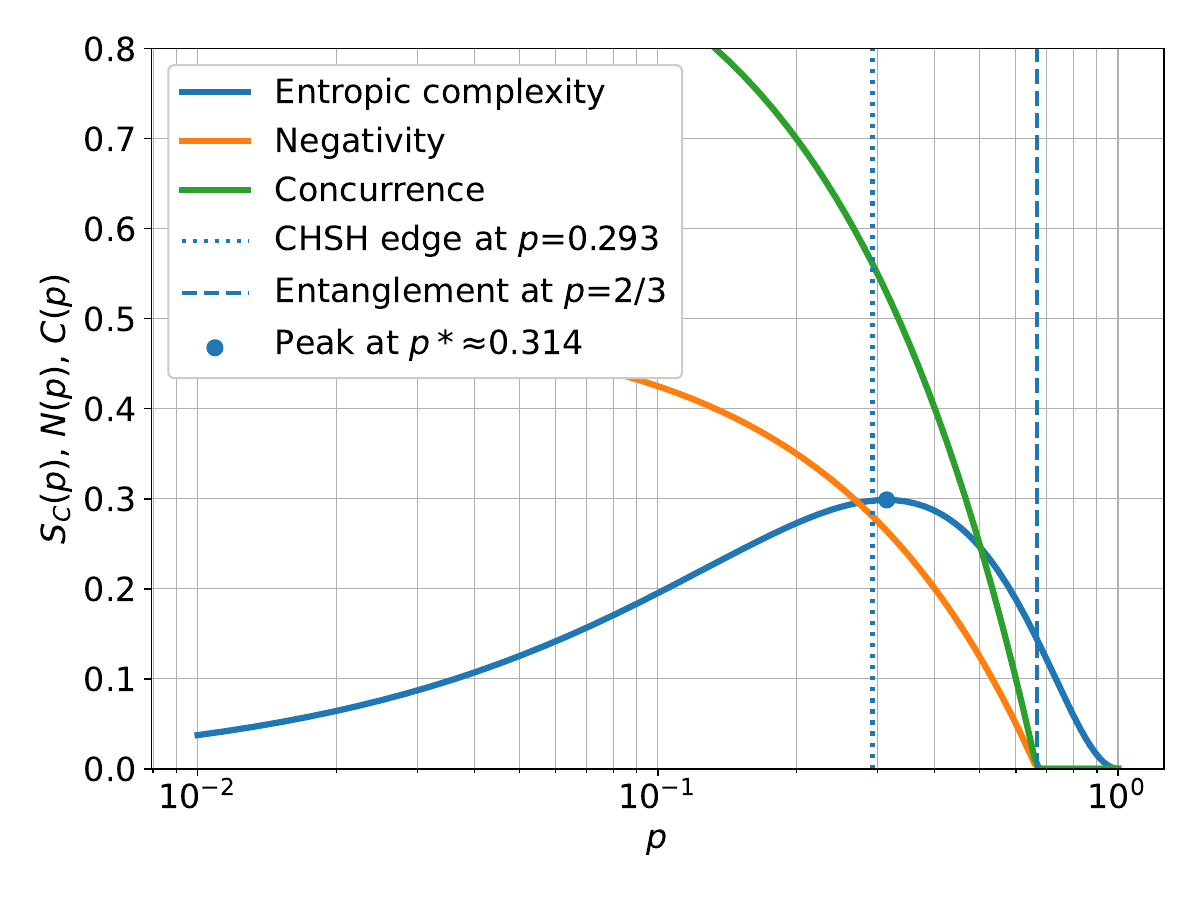}
\end{center}
\caption{Comparison of the $2$-qubit Werner state combined with a totally mixed state as given in Eq.~(\ref{werner_nsc}) as a function of parameter $p$ using several 
quantities, the entropic complexity, $S_C(p)$, the negativity, $N(p)$, and the concurrence, $C(p)$. 
The important limits of CHSH and entanglement edges are marked by vertical dotted and dashed lines.}
\label{fig:werner2}
\end{figure}
\begin{figure}
\begin{center}
\epsfxsize=8.4cm
\leavevmode
\epsffile{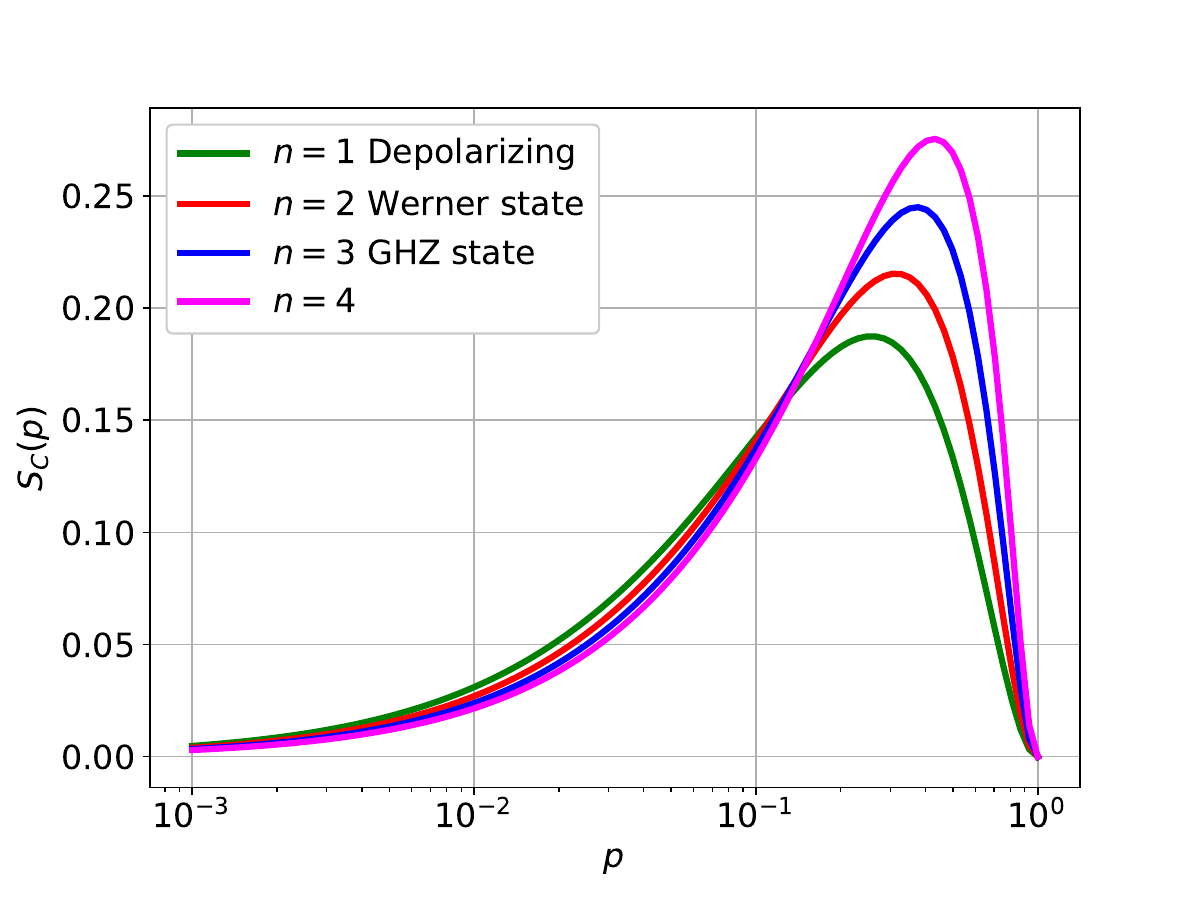}
\end{center}
\caption{Complexity of the $n$-qubit state combined with a totally mixed state as given in Eq.~(\ref{werner_nsc}) as a function of parameter $p$ for small values of $n$.}
\label{fig:werner_n1}
\end{figure}
\begin{figure}
\begin{center}
\epsfxsize=8.4cm
\leavevmode
\epsffile{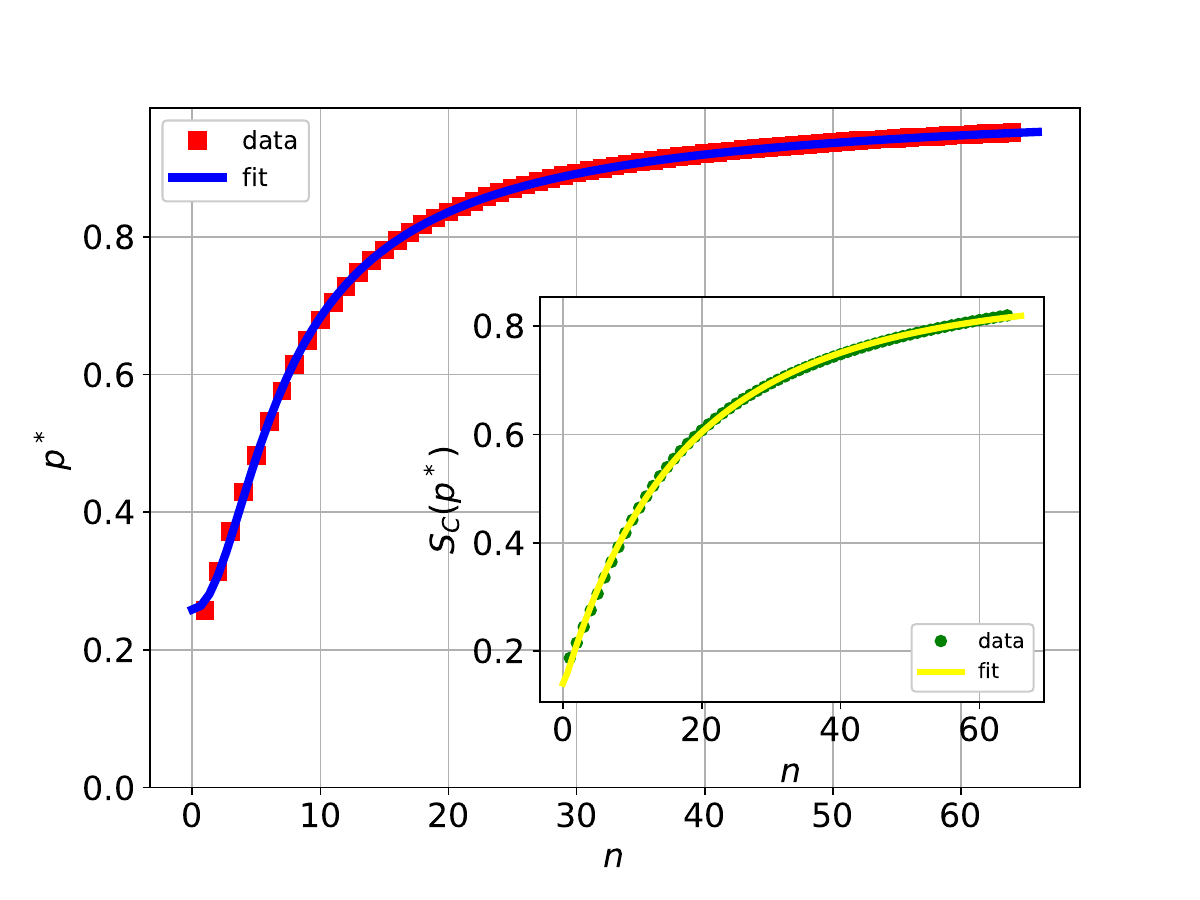}
\end{center}
\caption{The value of parameter $p^*$ with maximum complexity as a function of $n$, the number of qubits. The fit is approximately $1-p^*\sim n^{\gamma}$ with $\gamma\approx 1.05$
for $n\gg 1$. In the inset we plot the value of the maximum complexity at $p=p^*$ vs number of qubits, $n$. The fit is approximately $S_C(p^*)\sim n^{\delta}$ with $\delta\approx 0.14$.}
\label{fig:pmax_n}
\end{figure}
Below we define the interplay of the depolarization of the environment acting on a single qubit, then we extend the problem to the so-called Werner state, that is the combination
of a uniformly mixed state and a Bell-pair and finally we will investigate the case of an $n$-qubit generalization of this problem. Of course, the $n=3$ case is known as the GHZ~\cite{GHZ}
state. The entanglement properties are well-defined and investigated for $n=2$ and $n=3$, however, beyond purity, the entropic complexity can be applied even for any $n>3$.

\subsection{The depolarization channel of a single qubit}
As a first step one can take a single qubit even though there is no entanglement in that case but its quantum nature is essential. Hence in the 1-qubit limit the state to be combined
with an appropriate totally mixed state is
\beq
\rho (p) = (1-p)|\Phi^+\rangle\langle\Phi^+| + p \frac{\mathbb{I}}{2},
\label{werner_1}
\eeq 
if the qubit is represented as a symmetric combination of its ground and excited states as
\beq
|\Phi^{+}\rangle = \frac{1}{\sqrt{2}}\left(|0\rangle + |1\rangle\right).
\eeq
This case is well-known as the depolarization channel and has been investigated extensively~\cite{depolch}. It very well describes the problem that the quantum state is altered
by the environment with probability  $0\leq p\leq 1$ and remains unchanged with probability $1-p$. The eigenvalues of the density matrix are
\beq
\lambda_1(p)=1-\frac{p}{2} \qquad \lambda_2(p)=\frac{p}{2}.
\label{evalues1}
\eeq
Hence the entropic complexity becomes
\beq
\begin{split}
S_C(p)=-\frac{p}{2}\log\left(\frac{p}{2}\right) &- \left(1-\frac{p}{2}\right)\log\left(1-\frac{p}{2}\right) \\ 
                                                                              &+ \log\left(1-p+\frac{p^2}{2}\right).
\end{split}
\label{werner_1sc}
\eeq
These results show that the entropic complexity can be calculated for any quantum system, including a single qubit and hence can be generalized for a multi-qubit system.

\subsection{The 2-qubit Werner state}
The Werner state is an important 2-qubit state that interpolates between a Bell-state and a uniformly mixed state
\beq
\rho (p) = (1-p)|\Phi^+\rangle\langle\Phi^+| + p \frac{\mathbb{I}}{4}.
\label{werner_2}
\eeq
This is a prototype of a system with classical noise with  probability $0\leq p\leq 1$, while the system, represented by a Bell-state remains unchanged with probability $1-p$. 
The Bell-state can be either the symmetric or antisymmetric combinations of either both systems are in the ground state, $|0\rangle$ or the escited state, $|1\rangle$
\beq
|\Phi^{\pm}\rangle = \frac{1}{\sqrt{2}}\left(|00\rangle \pm |11\rangle\right),
\eeq
or similar combinations of either qubit is in the ground state and the other one is in the excited state
\beq
|\Psi^{\pm}\rangle = \frac{1}{\sqrt{2}}\left(|01\rangle \pm |10\rangle\right).
\eeq
In principle any of these Bell-states can be used but we will stick to the symmetric combination, $|\Phi^+\rangle$, as generalizations for other values of the number of qubits, $n$ is straightforward.

The eigenvalues of the density matrix as a function of the mixing parameter $p$ read as
\beq
\lambda_1(p)=1-\frac{3p}{4}, \qquad \lambda_2(p)=\lambda_3(p)=\lambda_4(p)=\frac{p}{4}
\label{evalues2}
\eeq
Therefore it is easy to calculate both the von Neumann and the R\'enyi entropies and the entropic complexity. Hence the latter reads as
\beq
\begin{split}
S_C(p) = &-\left(1-\frac{3p}{4}\right )\log\left(1-\frac{3p}{4}\right) - 3\frac{p}{4}\log\left(\frac{p}{4}\right) \\
                                                                                  &+\log\left(1-\frac{3}{2}p+\frac{3}{4}p^2\right).
\end{split}
\label{werner_2sc}
\eeq
This is the entropic complexity for any of the Bell-states. This entropy vanishes both for $p=0$ and $p=1$. Since both the von Neumann and the R\'enyi entropies have their maximum at $p=1$
with the value $S(1)=R_2(1)=\log(4)$. In the present case the 2 qubit system is 4 dimensional, hence this value appears in the above formulas. Furthermore, in order to keep normalized entropies,
we will always assume normalization with the maximum value, i.e. we will use the entropy pro single qubit. 

In Fig.~\ref{fig:werner2} we show the function given in Eq.~(\ref{werner_2sc}) together with the dependence of concurrence~\cite{Wootters}, $C(p)=1-3p/2$, and negativity~\cite{Horo}, $N(p)=(2-3p)/4$, 
as a function of $p$. The latter two are linear functions that vanish at the entanglement edge of $p=2/3$. Furthermore the so-called Clauser–Horne–Shimony–Holt (CHSH)~\cite{CHSH}  
limit is at $p=1-\sqrt{2}/2\approx 0.29289...$. The interval between these two limits is described as between the Bell separability and the entanglement edge. It is clear that the $S_C(p)$ curve attains
its maximum at a value, $p^{\ast}\approx 0.314$, that is between the CHSH limit and the entanglement edge.

\subsection{The n-qubit generalization}
Let us now generalize the above treatment of 2-qubit Werner state for 3-qubit or even for general $n$-qubit systems. 
In this case the dimensionality of the Hilbert-space is $d=2^n$, i.e. it grows exponentially as $n$ increases.
For instance the 3-qubit generalization of the state is nothing else but the so-called GHZ state~\cite{GHZ} and hence $d=8$.
\beq
|\Phi^{\pm}\rangle = \frac{1}{\sqrt{2}}\left(|000\rangle \pm |111\rangle\right).
\eeq
Without discussing the 3-qubit case separately, we take the general $n$-qubit case with $n\geq 2$. 
In that case the combined systems is given as
\beq
\rho (p) = (1-p)|\Phi_n^{\pm}\rangle\langle\Phi_n^{\pm}| + p \frac{\mathbb{I}}{d}.
\label{werner_n}
\eeq
where
\beq
|\Phi_n^{+}\rangle = \frac{1}{\sqrt{2}}\left(|0\rangle^{\otimes n} \pm |1\rangle^{\otimes n}\right).
\label{ent_state}
\eeq
The eigenvalues of this density matrix are simple. There is one prominent value
\beq
\lambda_1(p)=1-\frac{d-1}{d}p
\label{lambda1}
\eeq
and $d-1$-fold degenerate further eigenvalues
\beq
\lambda_k(p)=\frac{p}{d}, \qquad k=2\dots d-1.
\label{lambda2}
\eeq
One can easily check Eqs.~(\ref{evalues2}) and (\ref{evalues1}) for the special cases of $n=2, d=4$ and $n=1, d=2$.
Therefore the $n$-qubit generalisation of Eq.~(\ref{werner_2sc}) reads as
\beq
\begin{split}
S_C(p) = &-\left(1-\frac{d-1}{d}p\right)\log\left(1-\frac{d-1}{d}p\right) \\ &- (d-1)\frac{p}{d}\log\left(\frac{p}{d}\right) \\
                                                                                               &+ \log\left[1-\frac{2(d-1)}{d}p+\frac{d-1}{d}p^2\right].
\end{split}
\label{werner_nsc}
\eeq
In Fig.~\ref{fig:werner_n1} we plotted the entropic complexity, $S_C$ as a function of the mixing parameter $p$ for several values of the number of qubits involved, 
$n=1,2,3,4$. The prominent and remarkable property of the curves in Fig.~\ref{fig:werner_n1} is that besides vanishing for the extreme cases, $p=0$ and $p=1$,
as expected, they are simple and attain their maximum at a particular value of $p^{\ast}$. Therefore the system according to $S_C$ attains its maximal complexity at $p=p^{\ast}$, 
whose value increases as $n$ increases and seems to approach unity. At the same time the value $S_C(p^{\ast})$ increases as well. Apparently $p^{\ast}$ singles out 
that particular combination which still contains enough entanglement (quantumness) but already serve as a marking point at the quantum-classical crossover. That is the reason that for larger 
$n$ its value is getting closer to $p=1$, hence as $n$ increases the state with maximal complexity ought to have a major component from the completely mixed component and 
only smaller fraction from the fully entangled projection Eq.~(\ref{ent_state}). Needless to say that in the case of the $n$-qubit complexity $S_C(p)$ (\ref{werner_nsc}) can be analyzed 
analytically and for instance calculate its derivative
\beq
\begin{split}
\frac{dS_C}{dp}= & \frac{d-1}{d} \left [ \ln\frac{\lambda_1}{\lambda_2}-\frac{2(\lambda_1-\lambda_2)}{\lambda_1^2+(d-1)\lambda_2^2}\right ],
\end{split}
\eeq
where $\lambda_1(p)$ is the one in Eq.~(\ref{lambda1}) and $\lambda_2(p)=p/d$, also given in Eq.~(\ref{lambda2}).
Unfortunately this is a transcendental equation in order to find $p^{\ast}$ as a function of $n$ where the derivative vanishes. Here we will be satisfied with numerical approximate 
value of $p^{\ast}$. Therefore it is straightforward to find numerically the value of $p^{\ast}$ as a function of $n$ and also the value of complexity at that particular value: $S_C(p^{\ast})$. 

In Fig.~\ref{fig:pmax_n} we show $p^{\ast}$ as function of $n$ together with the best fit, that starts as a constant for $n\to 1$ and approaches the behavior $1-p^\ast\sim n^{-\gamma}$ 
behavior with $\gamma\approx 1.05$ for large enough $n\gg 1$. In the inset of Fig.~\ref{fig:pmax_n} we show $S_C(p=p^{\ast})$ as a function of $n$. Here we rescale this quantity, 
since both $S(p)$ and $R_2(p)$ have a trivial value at $p=1$ which is for both entropies equal, i.e. $S(1)=R_2(1)=\log(d)=n\log(2)$. For the other extreme, $p=0$, the entropies vanish, 
i.e. $S(0)=R_2(0)=0$. Therefore $S_C(p)$ for all cases, values of $n$ is rescaled by $\log(d)=n\log(2)$, so $S(1)=R_2(1)=1$ and $S_C(0)=S_C(1)=0$ plotting the entropies per qubit. 
The resulting entropies are shown as data points together with a best fit whose behavior is roughly linear for $n\to 1$ and it increases as $n^{\delta}$ with $\delta\approx 0.14$ for $n\gg 1$

\section{Multi-qubit state in noise induced phase damping}

\begin{figure}
\begin{center}
\epsfxsize=8.4cm
\leavevmode
\epsffile{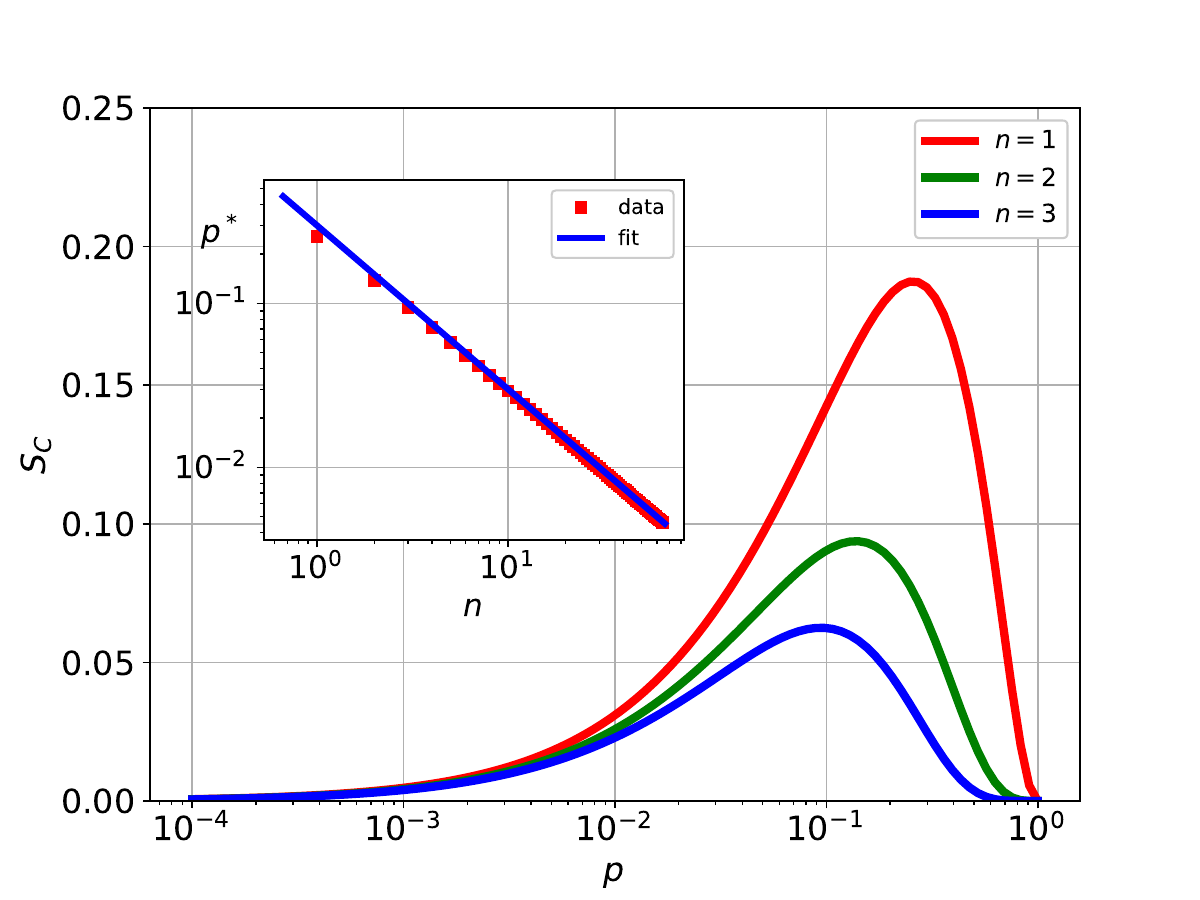}
\end{center}
\caption{The entropic complexity of the $n$-qubit system under the effect of dephasing with probability $p$. 
The inset shows the value of the maximum complexity as a function of the number of qubits, $n$ together with the best fit, $p^\ast \sim 1/n$.}
\label{fig:dephas}
\end{figure}

We now switch to a different analytically solvable noise model: local phase‑flip ($Z$) dephasing. Start from the $n$‑qubit GHZ state. 
Each qubit independently gets a Pauli‑$Z$ flip with probability $p$. The resulting state is 
\beq
\mathcal{E}(p)=pZ\rho Z + (1-p)\rho,
\eeq
where $Z$ is the Pauli-$Z$ matrix generalized for higher dimension $d=2^n$ and the original density matrix $\rho$ is based on the generalized, $n$-qubit version of the GHZ-state given in Eq.~(\ref{ent_state})
\beq
\rho = |\Phi_n^+\rangle\langle\Phi_n^+|. 
\eeq 
The density matrix perturbed by the external dephasing is simpler than the previous depolarization channel:
\beq
\begin{split}
\mathcal{E}(p)=& \frac{1}{2}\left ( |0\rangle\langle 0|^{\otimes n} +  |1\rangle\langle 1|^{\otimes n} \right ) \\ 
                         &+ \frac{1}{2}\gamma(p)\left ( |0\rangle\langle 1|^{\otimes n} +  |1\rangle\langle 0|^{\otimes n} \right ),
\end{split}
\eeq
where 
\beq
\gamma(p)=(1-p)^n
\eeq
There are only two nonzero eigenvalues of the density matrix
\beq
\lambda_{1,2}(p)=\frac{1}{2}\left[ 1\pm \gamma(p)\right ],
\eeq
meanwhile all the remaining, $d-2$ eigenvalues are zero. Hence it is straightforward to calculate the entropic complexity as
\beq
\begin{split}
S_C(p) = -\lambda_1(p)\log \lambda_1(p) & - \lambda_2(p)\log \lambda_2(p) \\ & + \log (\lambda_1^2(p)+\lambda_2^2(p)).
\end{split}
\eeq
So in terms of $\gamma(p)$
\beq
\begin{split}
S_C(p) = -\frac{\gamma(p)}{2}\log\frac{1+\gamma(p)}{1-\gamma(p)} + \frac{1}{2} \log \frac{(1+\gamma(p)^2)^2}{1-\gamma(p)^2}.
\end{split}
\eeq
As in the previous chapter one can also derive the derivative of this function versus $p$. It reads as
\beq
\begin{split}
\frac{dS_C}{dp} = -n(1-p)^{n-1}\left [ \frac{1}{2}\ln\frac{1-\gamma(p)}{1+\gamma(p)} + \frac{2\gamma(p)}{1+\gamma(p)^2} \right ]
\end{split}
\eeq
In order to find $p^{\ast}$ one has to solve this transcendental equation but we will be satisfied a numerical approximation.
The parameter dependence of the entropic complexity is shown in Fig.~\ref{fig:dephas} where the inset shows that the state with maximum complexity is the one where
$p^\ast\sim 1/n$.
The value of the entropic complexity at $p^\ast$ in this case is $S_C(p^\ast)= c_0 \log(d)$, with $c_0\approx 0.1875$ due to the simple structure of the eigenvalues of this density matrix.
\begin{figure*}
\begin{center}
\epsfxsize=16.8cm
\leavevmode
\epsffile{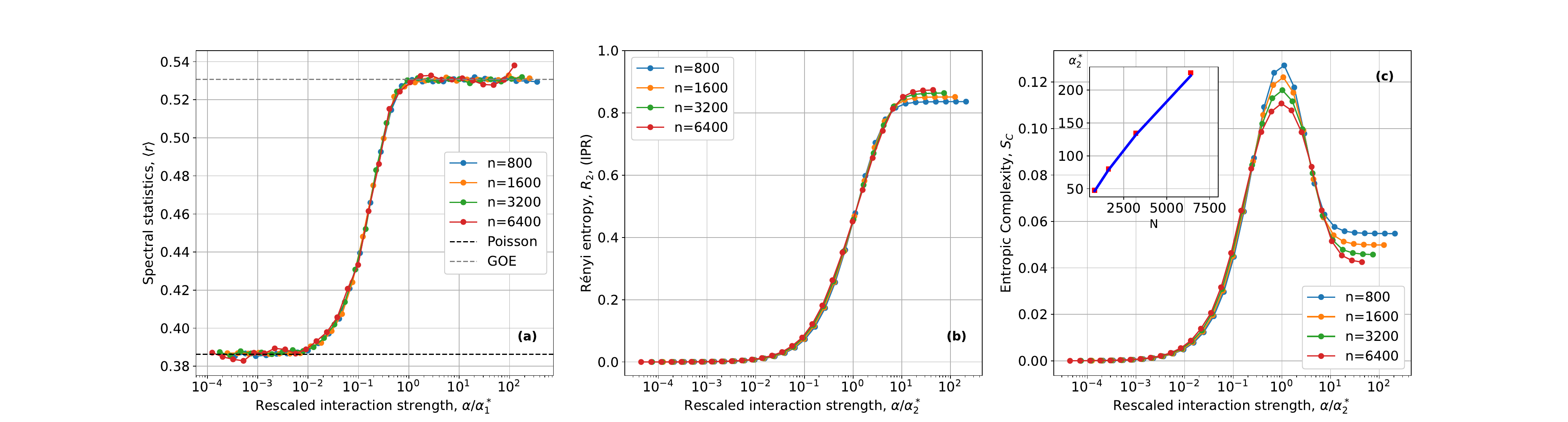}
\end{center}
\caption{Deformed GOE: (a) spectral statistics, (b) $R_2$, and (c) entropic complexity, $S_C$ for different $N$s.The scaling in (a) spectral statistics with respect to system size, $N$ 
is the known $\alpha_1^*~\sqrt{N}$. The scaling in (b) and (c) is $\alpha_2^*~N^{3/4}$. This scaling can be seen in the inset of (c) where the position of the peaks of $S_C$ is plotted as a function of $N$
together with a perfect power-law with the exponent of 0.75. In all cases the curves fall on top of each other.}
\label{fig:GOE}
\end{figure*}

\section{Complexity in Many-body Systems}

In this chapter we investigate systems described based on a Hamiltonian that resemble general many-body interactions. Here the solution is obtained by direct diagonalization and the states are analyzed in 
the so-called computational basis. The spin-spin or fermionic interactions are treated either based on random matrix techniques or using a simple model of many-body localization. Apparently any application of the
entropic complexity of the typical behavior of the eigenstates of these models shows a very good quantity in order to investigate smaller portions of the states of otherwise usually huge, exponentially large
Hilbert-space. 

Multi-qubit systems are fundamental building blocks of quantum information processing and quantum simulation. As the number of qubits increases, the complexity of their Hilbert space grows exponentially, 
giving rise to rich quantum many-body phenomena such as entanglement, thermalization, and quantum chaos. These systems serve not only as platforms for practical quantum computation but also as controlled 
environments for exploring deep questions in non-equilibrium statistical mechanics.

One particularly striking phenomenon that emerges in disordered interacting systems is many-body localization (MBL) — a phase in which ergodicity breaks down and the system fails to thermalize, even under its own unitary dynamics. 
In the MBL phase, local memory of initial conditions is preserved for arbitrarily long times, and entanglement spreads only logarithmically. This challenges the eigenstate thermalization hypothesis (ETH), which posits that individual 
eigenstates of chaotic systems appear thermal when probed locally.

In quantum simulation platforms such as trapped ions, Rydberg atoms, superconducting qubits, and ultracold atoms, multi-qubit dynamics can be engineered and measured with high precision, making them ideal for exploring 
the onset of MBL and transitions to ergodic phases. For instance, programmable spin chains realized on platforms like Google’s Sycamore or IBM’s Quantum Experience have enabled direct observation of MBL and its signatures,
~\cite{Roushan2017} including slow entanglement growth and persistent memory of initial states.

In second quantization, a quantum state is specified by listing how many particles occupy each mode (or site/orbital/spin). The Fock basis consists of all possible occupation number configurations.
Example (for fermions or spin-1/2 systems): Each mode (or site) can be either occupied (1) or unoccupied (0). The Fock basis state for a system with $L$ modes is written as:
\beq
|n_1, n_2, \dots, n_L\rangle
\eeq
where $n_i \in \{0, 1\}$ for fermions (due to the Pauli exclusion principle). For bosons, $n_i \in \mathbb{N}_0$. This basis arises from applying creation operators $a_i^\dagger$ to the vacuum state $|0\rangle$:
\beq
|n_1, n_2, \dots, n_L\rangle = \prod_{i=1}^L \frac{(a_i^\dagger)^{n_i}}{\sqrt{n_i!}} |0\rangle
\eeq
These basis states are eigenstates of the number operators $\hat{n}_i = a_i^\dagger a_i$.

Each mode in the Fock basis can be mapped to a qubit, especially in fermionic or spin-1/2 systems. The occupation number $n_i$ corresponds directly to the qubit state: $|0\rangle_i$ is an unoccupied mode corresponding to a qubit 
in state $|0\rangle$ whereas $|1\rangle_i$ is an occupied mode corresponding qubit in state $|1\rangle$. Thus, an entire many-body Fock state maps to a computational basis state of a multi-qubit system:
\beq
|n_1, n_2, \dots, n_L\rangle \quad \leftrightarrow \quad |n_1\rangle \otimes |n_2\rangle \otimes \dots \otimes |n_L\rangle.
\eeq
For example, the Fock state $|1, 0, 1\rangle$ maps to the 3-qubit state $|101\rangle$.

\subsection{Random Matrix Theory and Ergodicity Breaking}
\begin{figure*}
\begin{center}
\epsfxsize=16.8cm
\leavevmode
\epsffile{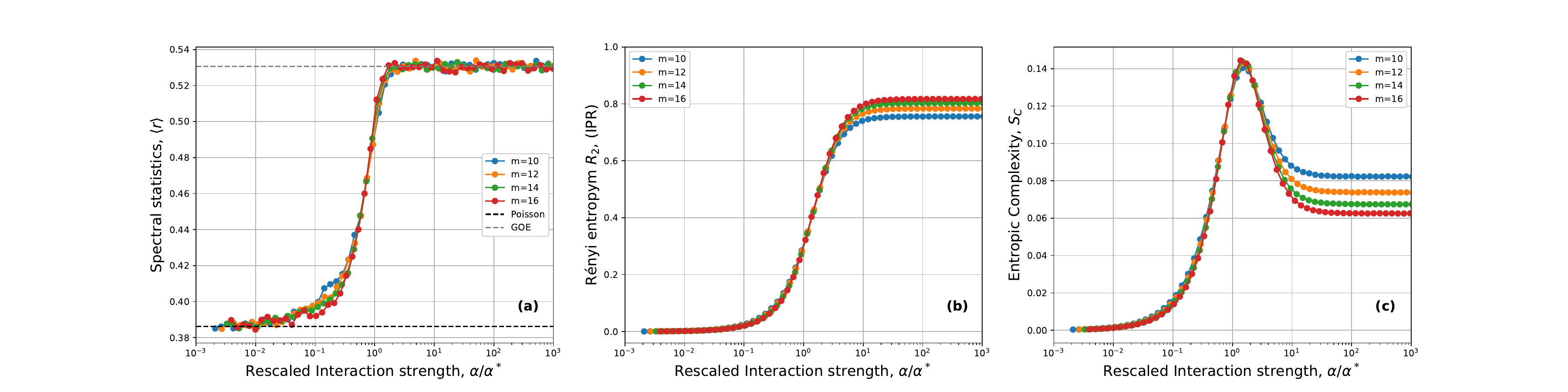}
\end{center}
\caption{Deformed TBRE: (a) spectral statistics, (b) $R_2$, and (c) entropic complexity, $S_C$ for different numbers of one-particle states, $m$ using $n=3$ as the number of fermions. The approximate scaling in $m$ and $n$ is
$\alpha^{\ast}\sim 1/n(m-n)$}
\label{fig:TBREm}
\end{figure*}
The statistical properties of many-body spectra provide powerful diagnostics for distinguishing ergodic and localized phases. In particular, random matrix theory (RMT) serves as a theoretical benchmark for chaotic many-body systems: 
ergodic systems exhibit level statistics that follow the Wigner-Dyson distribution (e.g., Gaussian Orthogonal Ensemble for time-reversal symmetric systems), characterized by level repulsion. In contrast, localized systems, 
lacking such level mixing, show Poissonian level statistics, signalling the absence of quantum chaos and the emergence of integrability through quasi-local integrals of motion.

RMT-based indicators — such as the average adjacent gap ratio $\langle r \rangle$ — have become standard tools to detect the MBL transition in both numerical simulations and experiments. These statistical methods complement 
dynamical signatures like the absence of thermalization, persistent spin imbalance, and slow entanglement growth. The first problem to study is the analysis of a combined random matrix model
\beq
H(\alpha)=H_0+\alpha H_1,
\label{deformed}
\eeq
where $H_0$ is an $N\times N$ random diagonal matrix and $H_1$ is an $N\times N$ member of the so-called GOE (Gaussian Orthogonal Ensemble), i.e. it is a real symmetric matrix. The parameter $N$ is the size 
of Hilbert-space and has only indirect relation to a multi-qubit system. Here by changing the parameter $\alpha$ the system undergoes an $N$ dependent crossover from a Poisson statistics for $\alpha=0$ to 
Wigner-Dyson statistics for $\alpha\to\infty$. In order to detect this transition the ratio $\tilde{r}_{\alpha}$ between neighbouring eigenlevels is defined as
\beq
\tilde{r}_{\alpha} = \min \left ( r_{\alpha}, \frac{1}{r_{\alpha}} \right),\qquad \text{where}\qquad r_{\alpha}=\frac{s_{\alpha}}{s_{\alpha -1}}
\eeq
and $s_{\alpha}=E_{\alpha+1}-E_{\alpha}$ is the spacing between neighboring levels. We know, that the average value over all eigenvalues for the Poisson distribution is $\langle \tilde{r}_{\alpha}\rangle\approx 0.39$ and for the
GOE, i.e. RMT case $\langle \tilde{r}_{\alpha}\rangle\approx 0.54$. In Fig.~\ref{fig:GOE}a we can see a clear, $N$ dependent transition between the integrable, Poisson and the chaotic, GOE statistics. Hence the crossover can 
be rescaled with  $\sqrt{N}$. Therefore, scaling of the spectral statistics is obtained with $\alpha_1^*\sim\sqrt{N}$.
On the other hand the statistical properties of the eigenstates show a remarkable parametrically different scaling as it is shown in Figs.~\ref{fig:GOE}b and \ref{fig:GOE}c. 
The behavior of the  average complexity of the eigenstates show a maximum whose position scales roughly as $\alpha_2^*\sim N^{3/4}$. Keeping the same rescaling on the behavior of the IPR (inverse participation ratio) 
in the form of the  R\'enyi entropy, $R_2$. 

In our deformed-GOE problem, the Shannon entropy, $S$ and the R\'enyi entropy are both computed in the $H_0$ basis. In the perturbative (Breit--Wigner) regime the spreading width is
$\Gamma \sim 2\pi\,\alpha^2\,\overline{|(H_1)_{ij}|^2}\,\rho_0 \sim \alpha^2/N$ since $(\rho_0 \sim 1)$, so the participation ratio scales as
$\sim \Gamma / \Delta_0 \sim \alpha^2/N$ since $\Delta_0\sim 1$. The entropic complexity typically peaks when eigenstates are highly mixed but not yet fully ergodic, 
i.e., in the fractal regime with correlation dimension (the scaling of IPR or $R_2$ with respect to $N$, $D_2\simeq 1/2$. Now equating $\alpha^2/N \sim N^{1/2}$ then gives the scaling of the position of the peak of $S_C$ as
$\alpha_2^*\propto N^{3/4}$. Hence in Figs.~\ref{fig:GOE}b and \ref{fig:GOE}c the rescaled variable is $\alpha/\alpha_2^*$. The inset of Fig. 5c shows the scaling of the peak position of $S_C$ which 
corresponds to the scaling of IPR obtained above from the perturbative arguments.
\begin{figure*}
\begin{center}
\epsfxsize=16.8cm
\leavevmode
\epsffile{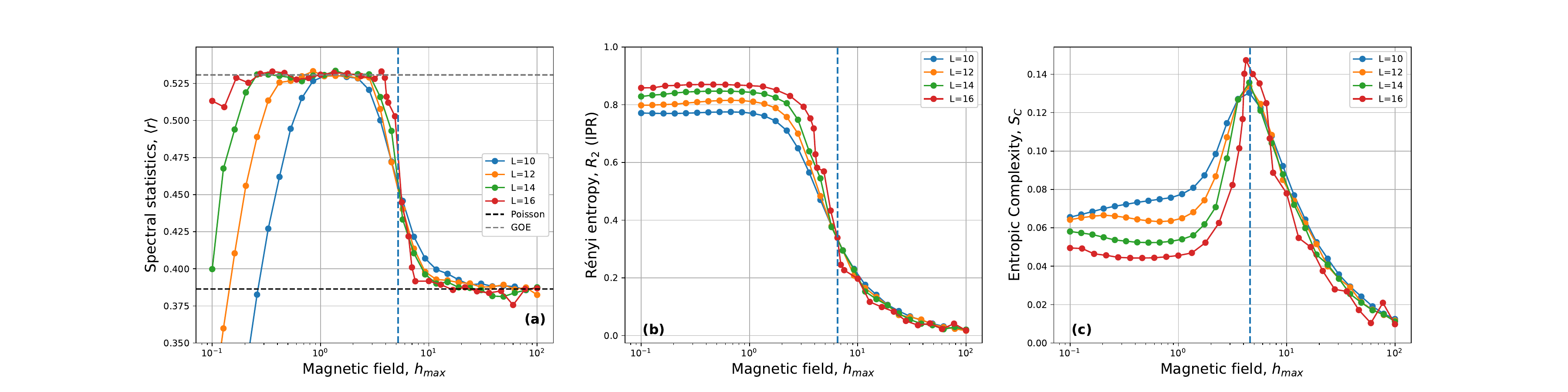}
\end{center}
\caption{MBL Transition: (a) spectral statistics, (b) $R_2$, and (c) entropic complexity as a function of the random magnetic field for different numbers of spins, $L$. In each subfigure the apparent scale independent approximant of
the critical magnetic field, $h_c$ is noted with a dashed line. For spectral statistics, $h_c\approx 5.5$, for $R_2$, $h_c\approx 6.0$, and for $S_C$, $h_c\approx 4.6$.}
\label{fig:MBL}
\end{figure*}

\subsection{Deformed Two-Body Random Interaction Ensemble}

In this subsection taking a step further we investigate the behavior of the spectral statistics and the eigenvector statistics of a deformed TBRE (Two-Body Random Interaction Ensemble)~\cite{Benet2003, TBRE, Jacquod2002, Aberg1990} 
similarly to the previous case
that is defined as in Eq.~(\ref{deformed}), where $H_0$ is an $N\times N$ diagonal random matrix, while $H_1$ is an $N\times N$ member of the TBRE. Here $N$ is defined on the basis of $n$ fermions distributed among 
$m$ one-particle states, hence $N = \binom{m}{n}$, so here $N$ grows exponentially with parameters $m$ and $n$. Note that for $n=2$ this model is identical to the previous one of full RMT problem with $N=m(m-1)/2$. 
This problem has been investigated using similar quantities in \cite{Jacquod2002} where it has been demonstrated that there exists a certain duality between $H_0$ and $H_1$ and the cross-over has been pointed out to depend 
parametrically different on $n$ and $m$ as opposed to the well-known criterion separating order from chaos based on spectral statistics~\cite{Aberg1990} which means that spectral fluctuations tend to reach the Wigner-Dyson limit 
for smaller interaction as opposed to the structure of the eigenstates. 

The TBRE is a more physically motivated model tailored for systems where only two-body interactions are relevant — a situation common in realistic quantum systems. It has a structured basis where the Hamiltonian includes 
random matrix elements only in the subspace of two-body interactions:
\beq
H(\alpha) = \sum_i \epsilon_i c_i^\dagger c_i + \alpha \sum_{i < j, k < l} V_{ijkl} c_i^\dagger c_j^\dagger c_k c_l,
\label{TBRE}
\eeq
where the diagonal part consists of a series of uncorrelated random energy values, $\epsilon_i$ chosen from a uniform distribution and the interaction parameters, $V_{ijkl}$ are drawn from a Gaussian distribution. 
The parameters of the random distributions are chosen such that the one-body spacing, $\Delta_1=\langle \epsilon_i-\epsilon_{i-1} \rangle$, is set to unity and the interaction matrix elements represent an appropriately 
sparse GOE matrix. The correlated sparsity is due 
to the two-body nature of the interaction where only those matrix elements between two basis vectors of Fock space are nonzero, which differ by at most two occupation numbers. 
This framework allows for the study of localization (as $\alpha\to 0$) in Fock space, the suppression of eigenstate thermalization, and the controlled transition between many-body ergodic 
($\alpha \langle V \rangle\gg\Delta_1$) and MBL-like behavior. 
The deformed TBRE allows further control over the interplay between single-particle structure and interaction-induced complexity. According to our present results corroborating our previous 
ones in~\cite{Jacquod2002} we see a markedly different parametric behavior in the scaling of the interactions strength, $\alpha$ with respect to the size of Fock space, $N$ in case of spectral 
and eigenstate statistics. 

As a result, Fig.~\ref{fig:TBREm} show the scaling of the data as a function of $\alpha$ keeping the number of particles, $n$ (Fig.~\ref{fig:TBREm}) fixed. 
In principle the $n$ and $m$ dependence is a lot more complicated as discussed in details in~\cite{TBRE}, but for the dilute limit of $m\gg n$ both scaling exponents are in good correspondence 
with the ones given in~\cite{TBRE}, $\alpha^{\ast}\sim 1/n(m-n)$. {We believe further investigation should clarify the details.}

\subsection{Many-body localization in 1D Heisenberg model}

Finally let us investigate a true many-body localization transition (MBL) that can be detected in a 1D Heisenberg model of spins subject to an external, random magnetic field~\cite{Pal2010}.
This model with disorder has become one of the most paradigmatic and widely studied models in the investigation of MBL,  a phenomenon in which interacting quantum systems fail to thermalize 
due to the presence of quenched disorder. In contrast to Anderson localization, which describes the absence of diffusion in non-interacting disordered systems, MBL represents a fundamentally richer 
regime where local interactions and disorder conspire to suppress thermalization, leading to emergent non-ergodic behavior in isolated many-body quantum systems, therefore it connects deeply 
with foundational questions in quantum statistical mechanics, such as thermalization, ergodicity breaking, and the validity of statistical ensembles in isolated quantum systems.

The problem can be described by the Hamiltonian
\beq
\hat{H} = J \sum_{i=1}^{L-1} \left( \hat{S}_i^x \hat{S}_{i+1}^x + \hat{S}_i^y \hat{S}_{i+1}^y + \hat{S}_i^z \hat{S}_{i+1}^z \right) + \sum_{i=1}^{L} h_i \hat{S}_i^z,
\eeq
where $J=1$ is set as a unit of energy, while $h_i$-s represent the effect of external, local, random magnetic field that is drawn from a symmetric, uniform distribution with zero mean and width $2 h_{max}$.
The MBL is expected to appear to be around $h_c\approx 4.5$ for small systems~\cite{Pal2010} of $L\leq 16$ using exact diagonalization, while more recent results over larger chains and scaling produce 
a critical value of $h_c\approx 3.7$ (see, e.g.~\cite{Luitz2015}).

Here the transition can be clearly depicted both for the average spacing ratio and the maximal point of the complexity measures of the eigenstates described by the entropic complexity, as you can see in
Fig.~\ref{fig:MBL}.  It is also remarkable, that the MBL can be located even at such small systems with $L=10,\dots 16$ spins. Note, the critical point 
strongly depends on the boundary conditions~\cite{Luitz2015, Pal2010, Zhang2022, Doggen2018}, therefore, we are not looking for exact values, but an estimate value of $h_c$ that already prominently shows up in the 
simulation of small systems. Similarly as in the previous cases the estimate for the critical magnetic field, $h_c$ for $\langle r\rangle$ is about $5.2\pm 0.5$, for $R_2$ it is $6.5\pm 0.5$ and for the $S_C$, it is
$4.6\pm 0.2$. Such a variation depending on observables has been shown in~\cite{Doggen2018}. {Note, however, the first and the last estimates overlap.} 
Clearly averaging over many more realizations and scaling with the length of the chain should be essential to produce more accurate data {that is beyond the scope of the present 
publication and will be the topic of further investigation.} Note, however, the Hilbert-space of $L$ spins even at the particular 
subspace where $S_z=0$ is very large, $D=L!/(L/2)!^2$, e.g. for $L=16$ the dimension of the subspace is $D=12.870$.

\section{Complexity of survival probability: Time Evolution in Chaotic Many-Body Systems}

Understanding the behavior of excited states in many-body quantum systems is a central challenge in modern physics, with implications ranging from fundamental statistical mechanics to quantum computation 
and condensed matter theory. In particular, multi-qubit systems — composed of interacting two-level systems — serve as natural platforms to explore quantum many-body dynamics, as they can capture both 
the complexity of interactions and the accessibility of computational modelling. 

In this context, the following question arises: How does a local or highly excited state evolve over time?~\cite{Flambaum2001} One way to probe this is through the survival probability defined as the probability that a system initially 
prepared in an excited many-body eigenstate (or product state) remains in or returns to that state after evolving under the full many-body Hamiltonian. In the language of multi-qubit system this problem can be formulated as
how long and in principle how can a multi-qubit state be preserved, what is the time scale that determines the 'melting' of the quantum information. As we will show here briefly and will work out in a subsequent publication
in details the entropic complexity, similarly as in the previous chapters, through its maximum marks a time scale of the destruction of the coherent state that depends on the interaction between the qubits on a nontrivial way. 

Mathematically, for an initial state $|\psi(0)\rangle$, the survival probability at time $t$ is given by:
\beq
P_{\text{ret}}(t) = |\langle \psi(0) | e^{-iHt} | \psi(0)\rangle|^2
\eeq
This function captures how quantum coherence and localization properties determine the spread or confinement of excitations. For thermalizing systems, the survival probability decays rapidly induced by dephasing as the excitation disperses 
over the full Hilbert space, and approaches the participation ratio in the energy basis, signaling ergodic dynamics and thermal behavior. In contrast, for localized systems — such as those exhibiting MBL — the decay can be arrested, 
indicating memory retention and non-ergodicity. Futhermore, late-time dynamics display correlation hole and special tails due induced by spectral edges.
   
In the following we will address the problem of an excitation in a system from the ground state and {using a toy model} we investigate the way this excitation is represented in the survival probability over a very large number of 
eigenstates.~\cite{Flambaum2001} Here a quantum state is represented as a many-body system of products of qubits. It is understood that there are altogether $N$ chaotic excited states with roughly equal nature, i.e. any excitation 
is thermalized over a very large set of states. 

Assume an initial state that is a basis state, $|0\rangle$ at $t=0$ and consider the time evolution of this state under the unitary evolution of the system. The weight of every single eigenstate, $W_i(t)$ obey the sum rule
\beq
W_0(t)+\sum_i^N W_i(t)=1,
\eeq
where in fact this represents the survival probability, i.e. $P_{\text{ret}}(t)=W_0(t)$. In case of thermalization, in our toy model we may {further} assume that $W_i(t)=W_f(t)$ for all $i=1\dots N$ is the same for all states, basically loosing 
the importance and the detailed information on the separate states that correspond to the quantum chaotic mixture. Therefore we have
\beq
W_f(t)=\frac{1}{N}[1-W_0(t)],
\eeq
so the actual form of $W_0(t)$ defines the whole dynamics. Using these quantities it is straightforward to calculate the von Neumann entropy as
\beq
\begin{split}
S(t)= & -\sum_i^N W_i(t)\log W_i(t) = \\
         & - W_0(t)\log W_0(t) - [1-W_0(t)]\log\left[\frac{1-W_0(t)}{N}\right ]
\end{split}
\eeq
and the R\'enyi entropy as
\beq
R_2(t)=-\log\sum_i^N W_i^2(t) = -\log\left [ W_0^2(t) + \frac{(1-W_0(t))^2}{N}  \right ].
\eeq
\twocolumngrid
Both of the above equations contain the embedding dimension of the Hilbert space, $N$. Usually instead of $N$ the participation number, $N_{pr}\leq N$ appears. 
Therefore the long time asymptotic is $R_2(t)\to\log N_{pr}$. The complexity quantity, $S_C(t)$ defined as
\beq
S_C(t)=S(t)-R_2(t),
\label{SCtime}
\eeq
behaves just like any other measures of complexity, since at $t=0$ and $t\to\infty$ it vanishes and being non-negative it attains its maximum at a certain value of $t$, that is a particular time scale
separating universal and non-universal behavior corresponding to the on-set of chaotic behavior. 

\begin{figure}
\begin{center}
\epsfxsize=8.4cm
\leavevmode
\epsffile{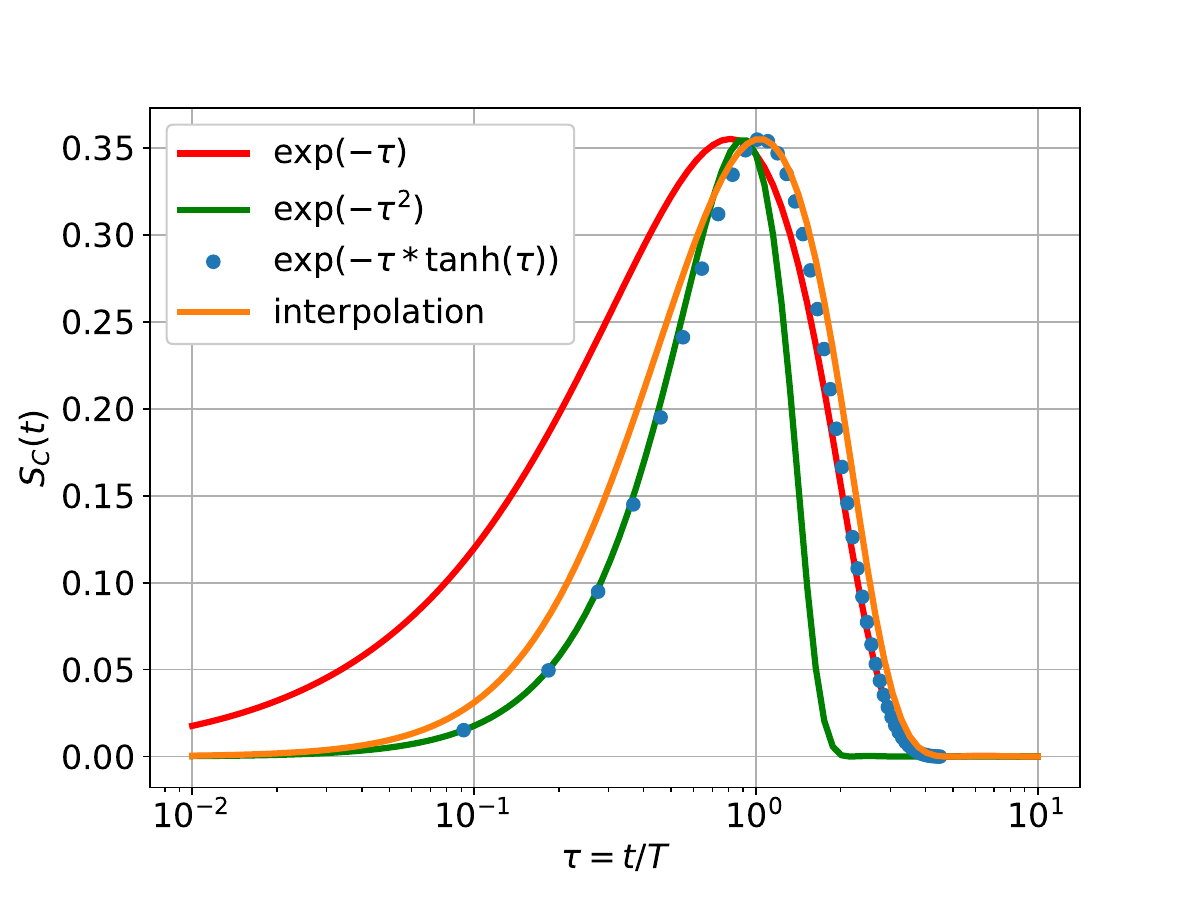}
\end{center}
\caption{Dynamical properties of the complexity of the survival probability for several simple decays and an interpolation formula as a function of $\tau=t/T$, with $T=1/\Gamma$.}
\label{fig:RetProb}
\end{figure}

Choosing typical forms of $W_0(t)$ we obtain different dynamical evolutions of $S_C(t)$. In Fig.~\ref{fig:RetProb} we depicted the behavior of $S_C(t)$ for several different survival probabilities of the
initial state, $W_0(t)$, one can clearly see the difference between its forms for exponential, i.e. $\exp(-t/T)$ or Gaussian, $\exp [-(t/T)^2]$ forms depending on $t/T$. Apparently several 
physical systems show a short time behavior, $t<T $, that is quadratic based on Fermi's golden rule originating in a Gaussian decay but for long times, i.e. beyond $t > T $ 
the exponential behavior maybe typical.  A trivial interpolating function between the exponential and the Gaussian can be obtained choosing $W_0(t)=t/T\tanh(t/T)$. However, a physically
more relevant case is when taking a two parameter interpolation~\cite{Flambaum2001}
\beq
W_0(t)=\exp\left ( \frac{\Gamma^2}{2\Delta^2} - \sqrt{\frac{\Gamma^4}{4\Delta^4} + (\Gamma t)^2}  \right),
\eeq
where $\Delta$ represents the width of the DOS (Density of States) and $\Gamma=1/\tau$ the rate of the long time decay of the excited states. 
In Fig.~\ref{fig:RetProb} we see the interpolation choosing $\Gamma^2/\Delta^2=2$.
 
In Fig.~\ref{fig:RetProb} it is clear that the complexity of this process is maximal at the time scale inherent in the system and the particular time dependence of the complexity depends deeply on the behavior 
of the way the survival probability behaves, which in turn is the Fourier transform of the density of states (DOS) of the system involved~\cite{Flambaum2001}. For instance a Gaussian density of states shows up 
in a Gaussian behavior of the survival probability meanwhile a Lorentzian DOS manifests in an exponential decay. The RMT semicircle DOS, on the other hand, produces a Bessel-function of the first kind.  
For the most part the excitation and hence the survival probability is usually monitored by calculating the increase in the von Neumann entropy as a function of time. Hereby we calculate the entropic 
complexity instead and show how it peaks at times represented by the inherent time scale, hence the position of this peak represents the inherent time scale dominating the problem. 

\begin{figure}
\begin{center}
\epsfxsize=8.4cm
\leavevmode
\epsffile{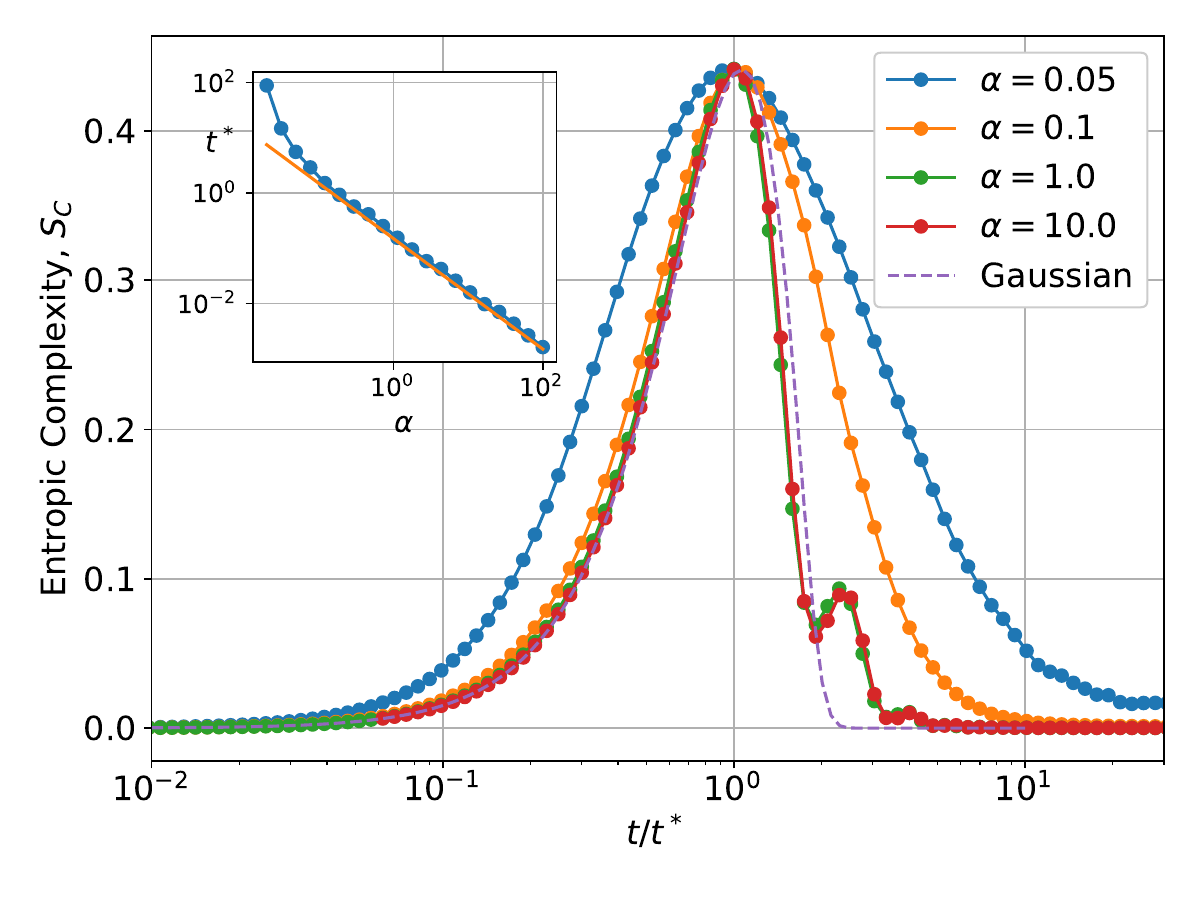}
\end{center}
\caption{Dynamics of the entropic complexity of the survival probability for the TBRE model as a function of $t/t^{\ast}$ for $m=14$ and $n=4$ for different values of the interaction parameter $\alpha$ together with the ideal case
of a Gaussian survival probability corresponding to a Gaussian density of states.
{The inset shows the  timescale, $t^{\ast}$ of the maximal value of $S_C$ as a function of the interaction parameter $\alpha$ for $n=4$ and $m=14$}}
\label{fig:TBREtime}
\end{figure}

In order to check the validity and applicability of this type of analysis we performed simple calculations on the TBRE model as described in Eq.~(\ref{TBRE}). Here a random configuration of qubits are affected by the 
two-body interaction term. One can determine the typical and average survival or return probability of any eigenstate of the unperturbed Hamiltonian, $H_0$. This is depicted in Fig.~\ref{fig:TBREtime}. The curves are labeled
by the interaction strength, $\alpha$. The inset shows the actual survival probability function, the main figure shows the time dependence of the entropic complexity, $S_C$ (see Eq.~(\ref{SCtime})) as a function of a rescaled time, 
$t/t^{*}$. The value of $t^{\ast}$ is given in the inset of Fig.~\ref{fig:TBREtime}. It is remarkable, that the timescale of the maximum of $S_C$ is independent of both the number of fermions, $n$, and the number of one-particle states, $m$
and depend only on the strength of interaction $\alpha$ presented as a universal behavior,  $t^{\ast}\sim 1/\alpha$ for large enough $\alpha$. Obviously a more detailed analysis is necessary to understand such behavior. 
In addition we should mention that the overall $\alpha$ dependence here corroborates what was found in~\cite{Jacquod2002} using the variation of the density of states with the increase of $\alpha$. A more detailed 
investigation reveals the duality that is inherent in the deformed TBRE problem between the role of $H_0$ and $H_1$ as pointed out in~\cite{Jacquod2002}.

\section{Conclusions}
Hereby we investigated multi-qubit states as a combination of their ground and excited states. With probability $p$ they have been changed by the decoherence due to the noise from the environment to a completely
mixed state while the original state is left unchanged with probability $1-p$. The characterization of the resulting combination, the mixed state is conducted by using an entropic complexity measure calculated from the difference
of the von Neumann and the R\'enyi entropy of order 2. The states with decoherence content $p\ll p^{\ast}$ are expected to be closer to the quantum, pure, entangled state, hence should be less vulnerable to the
effect of the noise while those states with $p\gg p^{\ast}$ are already close to being classical and thus more extended over the Hilbert-space. The cross-over value of $p^{\ast}$ is found to correspond to the case with
the complexity measure attaining its maximum that scales with the number of qubits, $n$. The asymptotic scaling of the form of $1-p^\ast\sim n^{-\gamma}$ with $\gamma\approx 1.05$  has been presented up to $n=63$ corresponding to
a Hilbert space of the size of $d=2^{64}\approx 10^{19}$. In case of a dephasing interaction with the environment, similar behavior was found with $p^\ast \sim 1/n$. Furthermore, we investigated the behavior of the entropic
complexity of many-body states using traditional deformed RMT and TBRE models, as well as the 1d Heisenberg-chain of spins under the action of local, random magnetic field exhibiting the MBL. It has been shown, 
that the cross-over and the transition between localized and ergodic many-body states is well described (beside traditional parameters, e.g. spectral ratio and IPR) by the entropic complexity measure. However, for these cases the 
ergodic states representing totally developed quantum chaos are described by a non zero complexity value and again whenever the entropic complexity is maximal a crossover or even a transition occurs. {Similar to a
recent publication~\cite{magic1}, apparently the magic or non-stabilizerness~\cite{magic2} detects the cross-over and transition in a similar way, eventhough magic is a kind of resource that is defined on different grounds. 
The relation or complementarity of our complexity will be addressed in a future investigation.}
In an investigation of the thermailzation process we have analyzed how the time scales and even the actual functional form the survival probability can be investigated in a closed system composed of a big number of states. We have
found that the increase of the internal interactions between qubits decreases the time scale beyond which quantum information is more likely to be lost substantially. This feature will be the focus of a subsequent publication.

In summary we have shown that the entropic complexity attains its maximum for states that represent a cross-over or a transition between pure quantum and pure classical nature. That particular state with maximal complexity still
shows enough quantumness for it to be considered to operate under the laws of quantum physics even in the presence of the effect of classical interactions but mark in parameter space a particular point which is essential in
the behavior of the system at hand. Therefore a quantum state with maximal entropic complexity is one that is highly entangled, delocalized, and statistically indistinguishable from a random state, yet not trivial — reflecting maximal 
internal quantum complexity within the allowed constraints of the Hilbert space. Maximal entropic complexity marks the edge of the operability in the parameter space of quantum devices. Beyond this point, quantum computation 
becomes infeasible, and the system behaves like a quantum thermal machine, not a computer, as it has been pointed out earlier, the quantum computer "melts"~\cite{Srednicki}.

\begin{acknowledgements}
Project no. TKP2021-NVA-02 has been implemented with the support provided by the Ministry of Culture and Innovation of Hungary from the National Research, 
Development and Innovation Fund, financed under the TKP2021-NVA funding scheme. Additional support was provided by the Ministry of Culture and Innovation and the 
NRDI Office within the Quantum Information National Laboratory of Hungary (Grant No. 2022-2.1.1-NL-2022-00004).
\end{acknowledgements}

\end{document}